\documentclass[traditabstract]{aa}
\usepackage{amssymb,amsmath,epsfig,bm,color,slashed,mathrsfs}
\usepackage{amsmath,bm}
\usepackage{txfonts}
\usepackage{natbib}
\usepackage{graphicx}
\usepackage{multirow}
\bibpunct{(}{)}{;}{a}{}{,}

\newcommand{\be}{\begin{equation}}
\newcommand{\ee}{\end{equation}}
\newcommand{\bea}{\begin{eqnarray}}
\newcommand{\eea}{\end{eqnarray}}

\newcommand{\ie}{{\it i.e., }}

\newcommand{\vecQ}{{\bm Q}}

\definecolor{red}{rgb}{0.8,0,0}
\definecolor{violet}{rgb}{0.4,0,0.4}
\definecolor{green}{rgb}{0,0.5,0.0}
\definecolor{navy}{rgb}{0.0,0.0,0.6}
\definecolor{orange}{rgb}{0.8,0.2,0.0}
\definecolor{blue}{rgb}{0.3,0.0,0.8}

\begin{document}

\title{
Rapid  cooling of the compact star in Cassiopea A \\as a  phase transition in dense QCD
}

\abstract{ 
  We present a model of the compact star in Cassiopea
  A that accounts for its unusually fast cooling behavior. This
  feature is interpreted as an enhancement in the neutrino
  emission triggered by a transition from a fully gapped, two-flavor,
  color-superconducting phase to a crystalline phase or an alternative
  gapless, color-superconducting phase. By fine-tuning a single
  parameter -- the temperature of this transition -- a specific
  cooling scenario can be selected that fits the Cas A data. Such a
  scenario requires a massive $M\sim 2M_{\odot}$
  star and is, therefore, distinctive from models invoking  canonical
 1.4 $M_{\odot}$  mass star with nucleonic pairing alone.  
}

\author{Armen Sedrakian}
\institute{Institute for Theoretical Physics, J. W. Goethe University,
D-60438  Frankfurt am Main, Germany}

\titlerunning{Rapid cooling of Cas A}

\date{\today}

\maketitle

\section{Introduction}
Recently, an unprecedented fast cooling of the compact star in
Cassiopeia A (Cas A) - the youngest known supernova remnant in Milky
Way - was inferred from the Chandra observations over a period of ten
years~\citep{2010ApJ...719L.167H,2011MNRAS.412L.108S}. The age of the
object 330~yr is tentatively deduced from the association of the
remnant with the  supernova SN 1680. This age estimate is
confirmed through kinematical estimates. The fits to the surface
temperature of the star suggest that the thermal soft X-ray spectrum
originates in a nonmagnetized carbon atmosphere at temperature
$\sim 2.\times 10^6$~K and emitting region of 8-17 km.

Hadronic models of compact stars have been invoked to fit the data by
assuming a canonical $1.4 M_{\odot}$ mass neutron star. The proposed
models include those that attribute the Cas A cooling to a star that
($a$) undergoes a transition to a superfluid state in the baryonic
core, which is accompanied by an enhancement of the cooling due to
Cooper pair-breaking
processes~\citep{2011PhRvL.106h1101P,2011MNRAS.412L.108S}.  These
models appear to be the most natural ones, since they do not require
``exotic'' physics, and the parameter values are within their accepted
range. In another model, ($b$), the star experiences thermal
relaxation in the core that is slower than commonly accepted owing to
low thermal conductivity~\citep{2012PhRvC..85b2802B}.  In addition,
the relation between surface and core temperature used in these
simulations is fitted to slowly cooling stars and deviates from the
relation used in the models in the previous item, ($a$).  An
alternative model suggests that the star ($c$) undergoes rotationally
induced compression that triggers a rapid Urca cooling in the center
of the star~\citep{2013PhLB..718.1176N}. This model requires a
significant fine tuning of the parameters and assumes an initial spin
of the star in the millisecond range, which has not been observed in
young neutron stars so far.

Recent observations of PSR J1614-2230 and PSR J0348+0432 lend firm
evidence that equilibrium sequences of neutron stars contain massive
$\sim 2 M_{\odot}$
members~\citep{2010Natur.467.1081D,2013arXiv1304.6875A}.  Models that
predict quark matter in such massive compact stars in
color-superconducting states have been studied
extensively~\citep{2005ApJ...629..969A,2007PhLB..654..170K,2008PhRvD..77b3004I,2009PhRvD..79h3007K,
  2010arXiv1006.4062K,2012A&A...539A..16B}.  Color superconductivity
is important;  otherwise, the unpaired quarks cool a
star by neutrino emission rapidly to temperatures well below those observed in Cas~A
\citep{2000PhRvL..85.2048P,2005PhRvD..71k4011A,2006PhRvD..74g4005A,2005PhRvC..71d5801G,2011PhRvD..84f3015H}.

The true ground state of QCD is not known at densities relevant for
compact stars. At relevant densities, the strong coupling
$\alpha_S\sim 1$, that is, the perturbation theory is not
useful. Furthermore, because of $\beta$-equilibrium and strange quark
mass, the Fermi-surfaces of up- and down-quarks are shifted apart.  As
a consequence, there are modifications to the standard
Bardeen-Cooper-Schrieffer (BCS) picture of pairing.  The non-BCS
pairing may lead to one of the possible gapless two-flavor
phases~\citep{2003PhLB..564..205S,2003PhRvD..67h5024M,2005JPhG...31S.849S}
or, for example, to the crystalline color-superconducting phase
\citep{2001PhRvD..63g4016A,
  2005PhLB..627...89C,2005PhLB..631...16G,2006PhRvD..74i4019R,2007PhRvD..75k4004G,2007PhRvD..76g4026M,2009PhRvD..80g4022S,2010PhRvD..82d5029H,2013arXiv1302.4264A}.
The crystalline-color-superconductivity comes in several variants,
which differ by the way the translational symmetry is broken by the
condensate of Cooper pairs carrying finite momentum.  Below we use
results obtained for the so-called Fulde-Ferrell phase (hereafter FF
phase), which is simple to model, but is general enough to preserve a
key feature of the crystalline phases, namely the existence of gapless
modes on the Fermi surfaces of
quarks~\citep{2001PhRvD..63g4016A,2009PhRvD..80g4022S,2010PhRvD..82d5029H}.

\begin{figure}[t]
\begin{center}
\includegraphics[width=8cm,height=7cm]{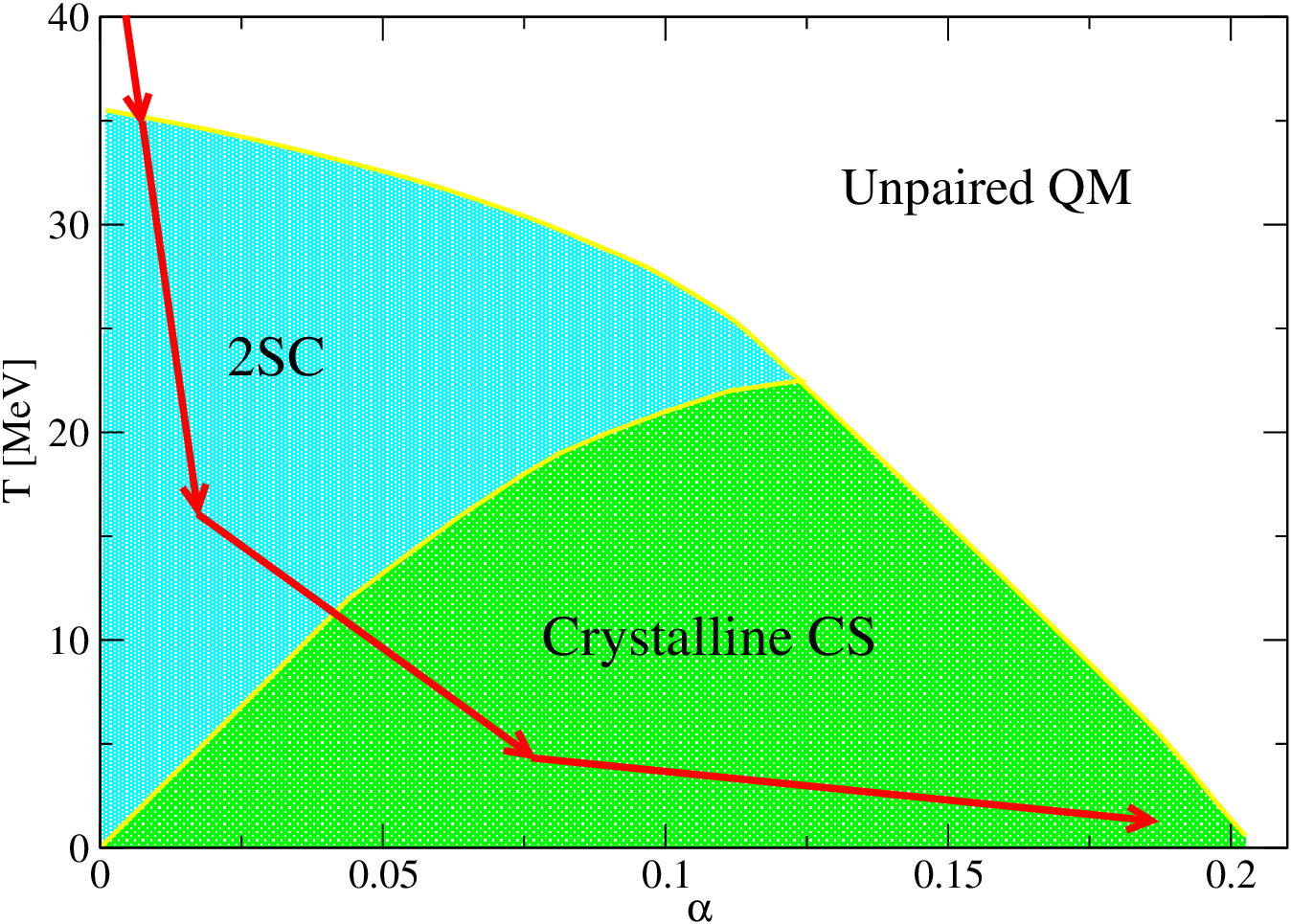}
\vskip 1cm
\includegraphics[width=8cm,height=7cm]{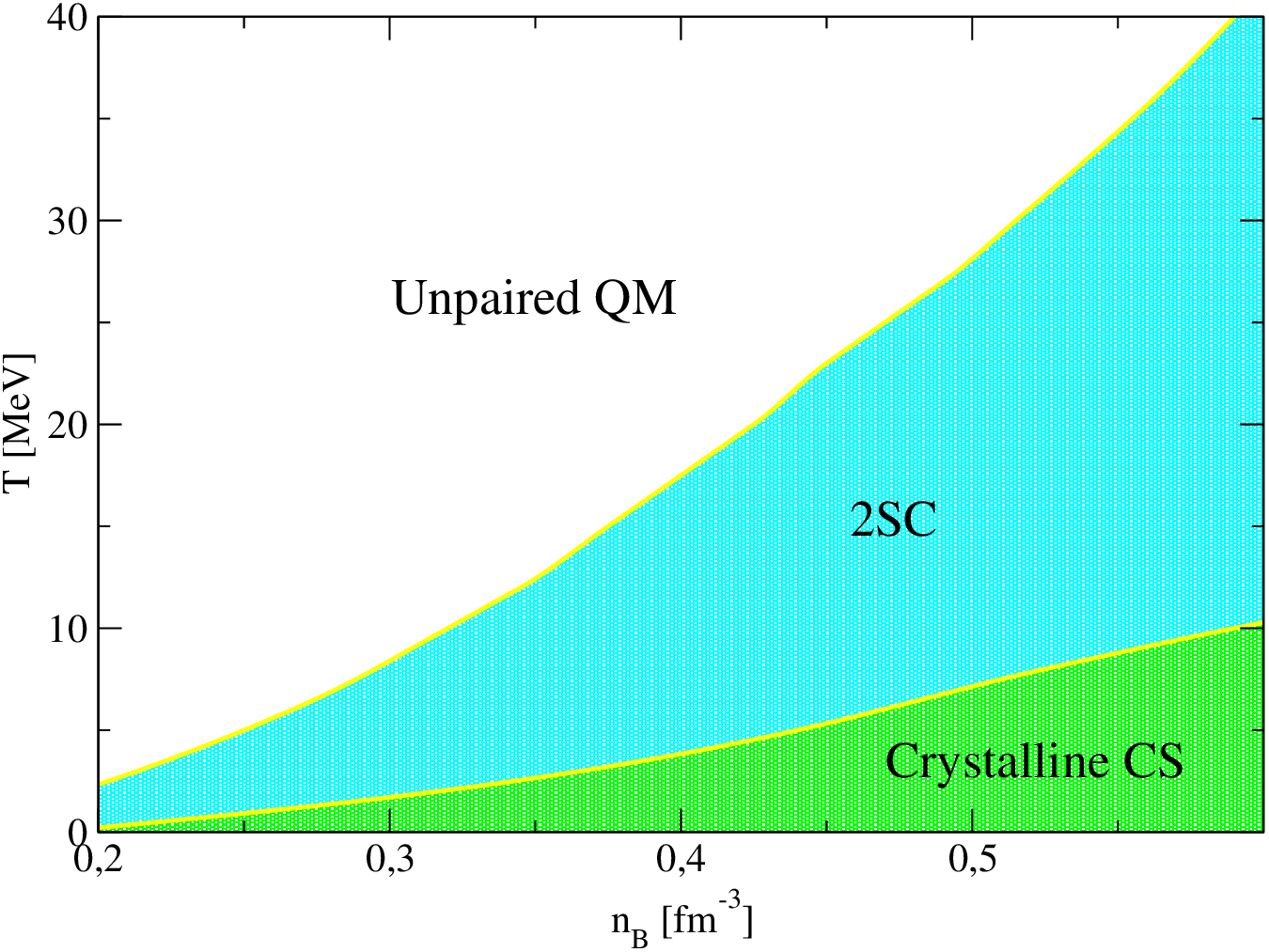}
\caption{{Upper panel:} Phase diagram of color-superconducting
  quark matter in the temperature ($T$) and isospin imbalance $\alpha
  = (n_d-n_u)/(n_u+n_d)$ plane, where $n_d$ and $n_u$ are the number
  densities of $d$ and $u$ flavors of quarks. The arrows 
  schematically show  the path in the phase diagram during the cooling of a
  star.  Lower panel: Phase diagram of the same phases, but in the
  temperature and baryon density ($n_B$) plane for matter in $\beta$-equilibrium
  with electrons.  }
\label{fig1}
\end{center}
\end{figure}

Proto-neutron stars emerge from supernova explosions with temperatures
$T\sim 50$ MeV and nearly isospin symmetric matter in the core.  After
the initial fast cooling to $T\sim 0.1$ MeV, the core reaches its
asymptotic state, which is highly isospin asymmetric. The initial fast
temperature drop is followed by much slower cooling to temperatures of
about 10 keV over the following $10^4-10^5$ years. Initially, the
high-temperature and low-isospin quark matter core is in the perfect
two-flavor, color-superconducting (hereafter 2SC) state where the
Fermi surfaces of the up- and down-quarks are completely gapped. As
the star cools and the $\beta$-equilibrium is established, the 2SC
phase becomes unstable, because the asymptotic ground state of quark
matter corresponds to the class of gapless and/or crystalline
two-flavor, color-superconducting phases mentioned above.  This is
illustrated in the top panel of Fig.~\ref{fig1}, where we show the
phase diagram of the FF and 2SC phases in the temperature and
flavor-asymmetry plane, adapted from
\cite{2009PhRvD..80g4022S}. 

During the early evolution of the star,
the matter traverses from the upper lefthand corner of the phase
diagram (i.e. from unpaired, almost isospin symmetric matter) to the
righthand lower corner, which corresponds to the cold FF phase and
large isospin asymmetry.  The phase diagram of the same phases in the
density-temperature plane and under $\beta$-equilibrium is shown in
the lower panel of Fig.~\ref{fig1}, which is adapted from
\cite{2010PhRvD..82d5029H}. These figures are only shown to illustrate
the structure and topology of the phase diagram. The magnitude of the
critical temperature of transition from the 2SC phase to the
crystalline color-superconducting or similar gapless phase will be
treated as a free parameter. Our fits to Cas A below show that it must be  
smaller than the values obtained from theoretical models.

In this work we conjecture that the rapid cooling of Cas A can be
understood as a phase transition from the perfect 2SC phase to a
crystalline/gapless, color-superconducting state. Thus, somewhat
counterintuitively, the rapid cooling is the result of a phase
transition {\it within} the phase diagram of dense QCD from one to
another color-superconducting phase.  This transition can be
characterized by two parameters, the critical transition temperature
$T^*$ and the transient ``widths'' $w$. The parameter $w$ encodes the
finite timescale of the transition.  In general $T^*$ will depend on a
number of factors, such as the nucleation history of the
superconducting phase, geometry of the superconducting regions,
dynamics and order of the phase transition, etc. Furthermore, because
of the density dependence of the critical temperature $T^*$ in the
stellar interior, this phase transition will not occur everywhere in
the quark core, but gradually within some shells. Therefore, the
values of $T^*$ and $w$ will be determined by the local conditions in
shells. We expect that the transition is not instantaneous, because it takes
place via  propagation of the phase-separation fronts rather than a
coherent instantaneous transition.  

The {\it equilibrium } model calculations give $T_c\sim 10$~MeV in a
particular (Nambu-Jona-Lasinio Lagrangian based) model, see
Fig.~\ref{fig1}. But the {\it absolute} value of the gap strongly
depends on the model. It may differ from the equilibrium critical
temperature of phase transition for dynamical reasons.  (Supercooled
state in electronic superconductor are observed, where the system
remains in the initial unpaired state below the equilibrium critical
transition temperature $T_c$, because of the finite timescales of the
phase transition process.)  These considerations motivate our
treatment of $T^*$ as a free parameter that will be used to fit our
cooling model to the Cas A data. The slope of the transient can be
chosen separately by the choice of the parameter $w$ prior to the fit
procedure. The simulations below will be carried out assuming a
density independent $T^*$. The density dependence seen in
Fig.~\ref{fig1} arises from the dependence of the density of states at
the Fermi-level alone; the coupling constants of the model are
density-independent. In general, the strong coupling constant of QCD
``runs''' with density (energy), therefore, the density dependence of
$T^*$ may be more complicated than in the model above. The possible
density dependence of $T^*$ will smooth out the transition to the
crystalline phase, which is taken into account phenomenologically in
the parameter $w$.

A preliminary report of the present study was given elsewhere
\citep{2013arXiv1301.2675S}.

\section{Modeling cooling processes}

Our cooling model is described in \cite{2011PhRvD..84f3015H}, where
the details of the microscopic input and the method of solution can be
found.  Here we describe the extension of this model, which allows us
to account for the phase transition between the phases in a
phenomenological manner.

Quark matter consisting of two light $u$ and $d$ flavors of quarks
cools predominantly via the beta decay (Urca) reactions $ d\to
u+e+\bar \nu,$ and $ u+e\to d + \nu, $ where $e$ stands for electron,
and $\nu$ and $\bar \nu$ represent electron neutrino and
antineutrino. For unpaired quarks and to leading order in the strong
coupling constant $\alpha_S$, the emissivity of the process {\it per
  quark color} was calculated by~\cite{1980PhRvL..44.1637I},
\be\label{eq:iwamoto} \epsilon_{\beta} =\frac{914}{945}G^2\cos^2
\theta p_d p_up_e\alpha_sT^6 , \ee where $G$ is the weak coupling
constant, $\theta$  the Cabibbo angle, and where $p_d$, $p_u$, and $p_e$ are
the Fermi momenta of down quarks, up quarks, and electrons.

Quark pairing modifies the temperature dependence of process
(\ref{eq:iwamoto}). In the BCS-type superconductors, the process is
suppressed almost linearly for $T\simeq T_c$ and exponentially for
$T\ll T_c$, where $T_c$ is the critical temperature.  In gapless
superconductors the emergence of the new scale $\delta\mu =
(\mu_d-\mu_u)/2$, where $\mu_{u,d}$ are the chemical potentials of
light quarks, leads to two qualitatively different possibilities,
which are distinguished by the value of the dimensionless parameter
$\zeta = \Delta_0/\delta\mu$, where $\Delta_0$ is the gap in the limit
$\delta \mu =0$~\citep{2006PhRvC..73d2801J}.  When $\zeta> 1$, the
entire Fermi surface is gapped. Consequently, when the thermal
smearing of the Fermi surface $\sim T$ is much smaller than the gap,
no excitation can be created out of the Fermi sphere. When $\zeta <
1$, particles can be excited from the gapless regions of the Fermi
sphere.  For the  FF phases the shift in the chemical
potential is replaced by the anti-symmetric in the flavor part of the
single particle spectrum of up- and down-quarks, $\epsilon_{u/d}$; \ie $\delta \mu \to
[\epsilon_d(\vecQ)-\epsilon_u(\vecQ)]/2$, where $\vecQ$ is the total
momentum of a Cooper pair. In the case of a single plane-wave
crystalline phase, the spectrum depends on the magnitude of the vector
$\vecQ$ alone.
\begin{figure*}[tb]
\begin{center}
\includegraphics[width=13.cm,height=7.5cm]{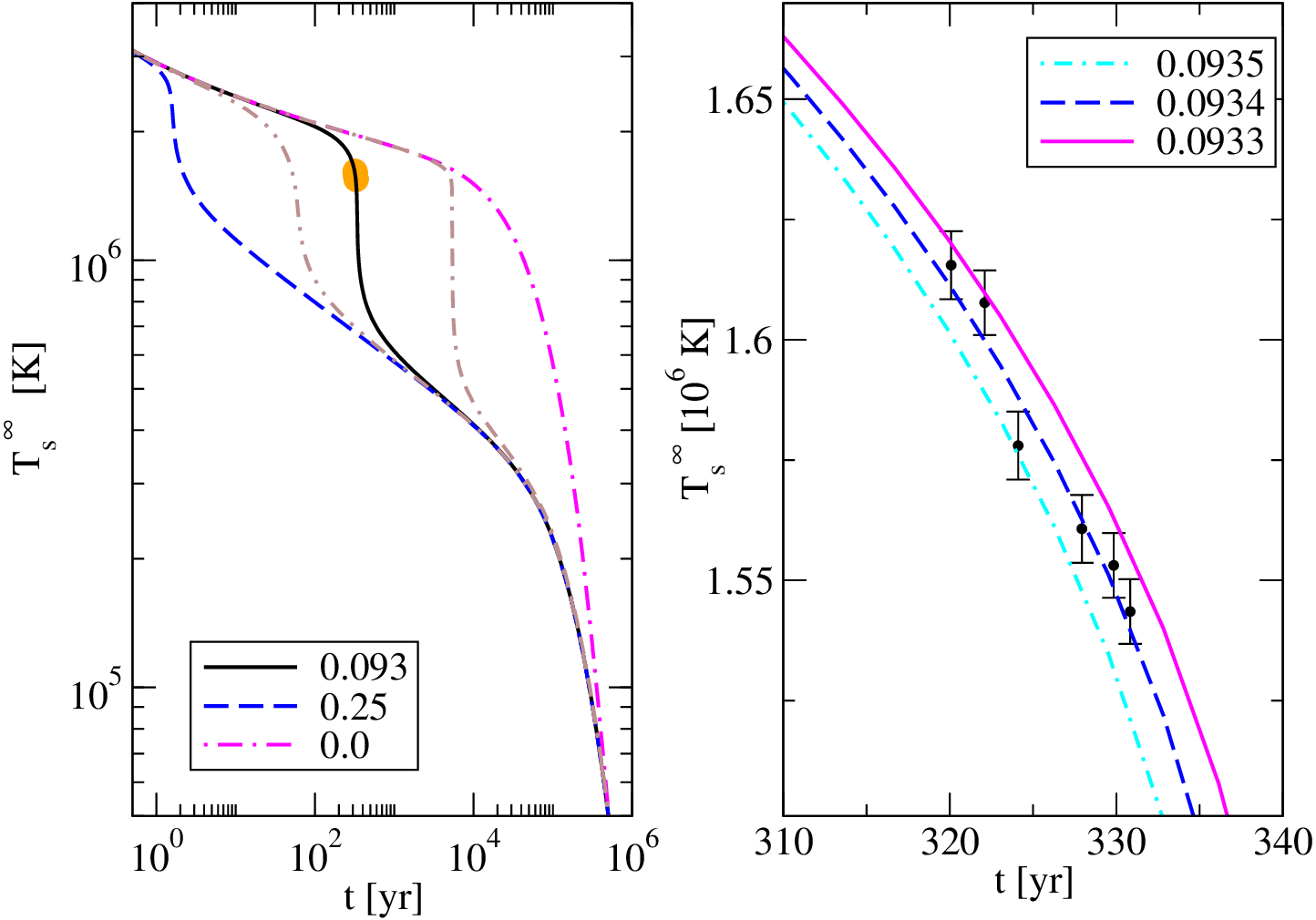}
\caption{{Left panel.} Dependence of the redshifted surface temperature
  (logarithmic scale) on time
  for the values of the transition temperature $T^*$ (in MeV)
  indicated in the plot and for fixed width $w = 0.2$ MeV. We also  show
  the same dependence for $T^* = 0.093$ and  $w = 0.3$ 
  (dash-double-dotted line) and $w= 0.1$ (double-dash-dotted line).  The
  underlying model is the 1.93 $M_{\odot}$ compact star model
  of~\cite{2009PhRvD..79h3007K}.  {Right panel.}  Dependence of the
  redshifted surface temperature (in units of $10^6$ K) on time for the same
  model.  The points with error bars show the Cas A data, and the solid
  lines are fits to these data by variation of $T^*$  for fixed
  width $w=0.2$ MeV.  The redshifted temperature measured by a distant
observer is obtained as $T_S^{\infty} = T_S e^{\phi_S}$, where
$\phi_S$ is the surface value of the function $\phi$, 
which is defined via the temporal component of the spherically
symmetrical metric according to $g_{00} =e^{2\phi}$.
}
\label{fig2}
\end{center}
\end{figure*}
The asymptotic behavior of the $\zeta$ parameter discussed above
permits us to use an interpolation formula that covers the
asymptotic limits and does not require detailed modeling of the
physics of the underlying phase~\citep{2006PhRvC..73d2801J}
\bea
\epsilon^{rg}_{\beta}(\zeta;T\le T_c) = 2f(\zeta) \epsilon_{\beta},
\quad f(\zeta) = \frac{1}{\exp[(\zeta-1)\frac{\delta\mu}{T}-1]+1},
\eea 
where the factor two is the number of colors.  

 Because the 2SC pairing pattern breaks the SU(3)
  color symmetry, one of the quark colors (say, the blue color) is not
  involved in that pairing. The strength and the flavor content of
  pairing among blue quarks is model
  dependent~\citep{2003PhRvD..67e4018A,2003PhRvL..91x2301S,2005PhRvD..71e4016S,
    2005PhRvD..72c4008A,2006PhRvD..74k4005A,2007Ap&SS.308..443A}.  It
  is likely that cross-flavor (blue-up and blue-down) pairing is
  suppressed by a large mismatch between the Fermi surfaces of the up
  and down quarks.  Pairing is expected in the color ${\bm 6}_S$ and
  flavor ${\bm 3}_S$ channel, which is same-flavor and same-color
  pairing, so is not affected by the flavor
  asymmetry~\citep{2003PhRvD..67e4018A}. However, the pairing in this
  channel vanishes in the chiral limit, when the mass of blue quarks
  vanishes. Pairing due to alternative mechanisms is not excluded, besides
those  studied so far.  Quite generally, single-flavor
  and/or color pairings have gaps in the range of 10-100 keV, which
  are much smaller than the gaps emerging from 2SC-type pairing.
  Since there are no definitive theoretical predictions about the
  pairing in the blue sector, we assume that the neutrino
  emissivity of blue quarks is suppressed exponentially by the pairing
  gap, as is the case for nucleonic matter.  Therefore, we
  simply parameterize the modifications of the Urca process on blue
  quarks as in the case of baryonic matter: 
\be
  \epsilon^{b}_{\beta}(T\le T_c)
  =\epsilon_{\beta}^b\exp\left(-\frac{\Delta_b}{T}\right), \ee where
  $\Delta_b$ is the pairing gap for the blue color quarks.  Note that
  $\epsilon_{\beta}^b = \epsilon_{\beta}^{rg}/2$ owing to the different
  numbers of colors involved. Several studies used similar
  parameterizations~\citep{2005PhRvC..71d5801G,2011PhRvD..84f3015H}.
  Furthermore, we  choose to work with a pairing gap $\Delta_b =
  100$ keV, which leads to blue quarks not
  affecting the cooling dynamics or the fits to Cas A data; \ie they
  only play a passive role.  The effect of lower values of this gap
  on the cooling is discussed elsewhere~\citep{2011PhRvD..84f3015H}.

  In addition to neutrino emissivities, we need to model the specific
  heat of the quark phases.  For the inhomogeneous red-green
  condensate, the critical temperature changes with
  $\delta\mu$~\citep{2011PhRvD..84f3015H} \be T_c(\zeta) \simeq
  T_{c0}\sqrt{1-\frac{4\mu}{3\Delta_0} \delta (\zeta)}, \ee where $\mu
  =(\mu_d+\mu_u)/2$, $\Delta_0 = \Delta(\zeta=0)$, $T_{c0} =
  T_c(\zeta=0)$, and $\delta (\zeta) = (n_d-n_u)/(n_d+n_u)$.  The
  fully gapped and gapless regimes behave differently; in the presence
  of gapless modes, the specific heat has a linear dependence on the
  temperature. A phenomenological way to model this behavior is given
  by the relation 
\bea 
c_S^{rg}(\zeta;T\le T_c) &=& f(\zeta) c_N^{rg},
 \eea 
  where $c_N^{rg}$ is the specific heat of red-green unpaired
  quarks and $c_S^{rg}$ is the specific heat of pair-correlated
  quarks.  To compute the specific heats of the electrons and unpaired
  quarks, we assume that each form noninteracting, ultrarelativistic
  gas.

  Finally, we need the relation between the temperature of the
  isothermal interior and the surface, which can be written as
  \citep{1983ApJ...272..286G,1997A&A...323..415P} \be T_{s6} = (\alpha
  T_9)^{\beta} g_s^{1/4} \ee where $T_{s6}$ is the surface temperature
  in units of $10^6$ K, $T_9$ the isothermal core temperature in units
  of $10^9$ K, $\alpha$ and $\beta$ are constants that depend on the
  composition of the star's atmosphere, $g_s$ is the surface gravity
  in units of $10^{14}$ cm s$^{-2}$.  For the present model with
  $M/M_{\odot} = 1.93$ and $R = 13.32$ km we find $g_s =1.91$.  We
  assume $\beta= 0.55$, which lies between the values for the
  purely-iron ($\beta = 0.5$) and fully accreted ($\beta = 0.61$)
  envelope.  We adopt the value $\alpha = 18.1$, which is appropriate
  for a accreted envelope~\citep{1997A&A...323..415P}.  The fit
  results are sensitive to the $\beta$ parameter, but not to the
  $\alpha$ parameter.

\section{Simulation results}

The key idea of our model is that the transition occurs from the fully
gapped (perfect) 2SC phase to some type of crystalline/gapless phase. A
particular realization of this scenario would be a transition to
single plane wave crystalline phase (the FF phase). We model this transition
phenomenologically by adopting temperature dependent (and, therefore,
time-dependent) parameter $\zeta (T)$. (\cite{2011PhRvD..84f3015H}
assume constant $\zeta$, so  phase transitions of the type discussed
here were excluded in their study from the outset). The temperature 
dependence of this parameter  is modeled such
that one obtains $\zeta > 1$ for $T> T^*$, whereas 
$\zeta < 1$ for $T< T^*$. The $\zeta(T)$ function is modeled by the simply law 
\be
\zeta(T) = \zeta_i - \Delta\zeta ~g(T), \ee where $\zeta_i$ is the
initial value, $\Delta\zeta$  the constant change in this function,
and the function $g(T)$ affects the transition from the initial value
$\zeta_i$ to the asymptotic final value $\zeta_f = \zeta_i -
\delta\zeta$. The transition is conveniently
modeled by the following function \be g(T) =
\frac{1}{\exp\left(\frac{T-T^*}{w}\right)+1}, \ee which allows us to
control the temperature  of transition, fixed by $T^*$, and the smoothness
of the transition, fixed by  width $w$.

The cooling simulations were carried out using the model of
\cite{2011PhRvD..84f3015H}, which solves the general relativistic
equations of cooling in spherical symmetry (see also
\cite{1980ApJ...239..671G}). The time evolution is carried out for the
quantity $\tilde T = Te^{\phi}$, where $T$ is the local temperature
and $\phi$ is related to the temporal component of the metric tensor
as $g_{00} = e^{2\phi}$.

The lefthand panel of Fig.~\ref{fig2} shows the cooling evolution of
$M = 1.93 M_{\odot}$ model of a compact star for several values of
$T^*$ and fixed $w = 0.2$ MeV. In addition, the figure shows the
effect of keeping $T^*$ constant and varying of $w$. Evidently the
drop in the temperature is smoother in the case of larger $w$. Once a
suitable value of $w$, which fits the slope of the transient, is
assumed, one can proceed to fit the data by choosing the temperature
$T^*$. It is seen that for a specific value of $T^*$ the cooling curve
passes through the location of Cas A in the log $T$-log $t$ plane.
The righthand panel of Fig.~\ref{fig2} shows the observed evolution of
Cas A for the same value of $w$ and two (fine-tuned) values of
$T^*$. The numerical values of gaps in the red-green and blue channels
$\Delta_{0} = 60$ MeV and $\Delta_{b} = 0.15$ MeV were used in the
simulations. The critical temperature of the
same-color pairing of the blue quarks satisfies the condition $T_{cb}
= \Delta_b/1.76 >T^*$, which implies that blue quarks have no effect
on the dynamics of the transition process.  In case of low
critical temperatures of blue quark pairing, a compact star will
rapidly cool via neutrino emission on time scales $t\le 300$ yr below the
temperature of Cas A, which obviously will preclude any fit to the
data. 

In closing, it should be stressed that the
  data can be fitted by choosing a range of combinations of $w$ and
  $T^*$ parameters, the first of which encodes unknown physics of
  dynamics of the phase transition and the density dependence of
  critical temperature. Therefore, even though some fine tuning of
  $T^*$ is required to fit the data, the actual range of $T^*$ by
  which the data can be fitted is in fact much broader. Better
  understanding of these factors, as well as of the magnitude of the
  critical temperature, are required in order to constraint the range
  of these parameters.

\section{Conclusions}

We have demonstrated that the rapid cooling of Cas A can be understood
as a phase transition from the perfect two-flavor,
color-superconducting phase to a gapless/crystalline
phase. Counterintuitively, {\it the phase transition occurs within the
  phase diagram of dense QCD and reflects the ordering of various
  superconducting phases in the temperature, density, and isospin
  spaces.}  This idea is supported qualitatively by the structure of
the phase diagram of the dense QCD at densities and temperatures
relevant to the cooling evolution of compact stars.

We have shown that the data can be fitted by adjusting a single
parameter - the phase transition temperature - at a fixed value of 
transient time scale, which encodes the dynamics of the transition.  A
further important assumption is that there is no interference from the
blue quarks, so that the temperature of the model is above the Cas A
temperature for $t\le 300$ yr, which is an obvious requirement for any
successful fit to data in the absence of internal heating.

The present model implies that only massive compact stars undergo the
rapid temperature drop observed in Cas A, for only massive members of
compact star sequences contain (color-superconducting) quark matter in
their interiors. It also implies that massive compact stars that are
older than Cas A are invisible in X-rays with modern instruments.  Our
model thus differs from the alternatives, which invoke nucleonic
pairing and pair-breaking
processes~\citep{2011PhRvL.106h1101P,2011MNRAS.412L.108S} or slow
thermal relaxation~\citep{2012PhRvC..85b2802B} and also work for 
low-mass mass neutron stars. Rotation is unimportant for our modeling as
opposed to the model of \cite{2013PhLB..718.1176N}.  It is evident that the
various models can be distinguished through their distinctive features
in case further observational information becomes available.
Clearly, it would be premature to conclude that the Cas A behavior is
{\it direct evidence of superfluidity for any particular type of
  superfluid} in neutron star interiors.

\section*{Acknowledgments}
The author is grateful to M. Alford, H.~Grigorian, D. Hess,
X.-G.~Huang, K.~Rajagopal, D.~H.~Rischke, A. Schmitt, and I.~Wasserman
for useful discussions and correspondence. This work is partially
supported though a collaborative research grant of the Volkswagen
Foundation (Hannover, Germany).

\bibliographystyle{aa} 

\begin{thebibliography}{42}
\expandafter\ifx\csname natexlab\endcsname\relax\def\natexlab#1{#1}\fi

\bibitem[{{Aguilera}(2007)}]{2007Ap&SS.308..443A}
{Aguilera}, D.~N. 2007, \apss, 308, 443

\bibitem[{{Aguilera} {et~al.}(2005){Aguilera}, {Blaschke}, {Buballa}, \&
  {Yudichev}}]{2005PhRvD..72c4008A}
{Aguilera}, D.~N., {Blaschke}, D., {Buballa}, M., \& {Yudichev}, V.~L. 2005,
  \prd, 72, 034008

\bibitem[{{Aguilera} {et~al.}(2006){Aguilera}, {Blaschke}, {Grigorian}, \&
  {Scoccola}}]{2006PhRvD..74k4005A}
{Aguilera}, D.~N., {Blaschke}, D., {Grigorian}, H., \& {Scoccola}, N.~N. 2006,
  \prd, 74, 114005

\bibitem[{{Alford} {et~al.}(2001){Alford}, {Bowers}, \&
  {Rajagopal}}]{2001PhRvD..63g4016A}
{Alford}, M., {Bowers}, J.~A., \& {Rajagopal}, K. 2001, \prd, 63, 074016

\bibitem[{{Alford} {et~al.}(2005{\natexlab{a}}){Alford}, {Braby}, {Paris}, \&
  {Reddy}}]{2005ApJ...629..969A}
{Alford}, M., {Braby}, M., {Paris}, M., \& {Reddy}, S. 2005{\natexlab{a}},
  \apj, 629, 969

\bibitem[{{Alford} {et~al.}(2005{\natexlab{b}}){Alford}, {Jotwani}, {Kouvaris},
  {Kundu}, \& {Rajagopal}}]{2005PhRvD..71k4011A}
{Alford}, M., {Jotwani}, P., {Kouvaris}, C., {Kundu}, J., \& {Rajagopal}, K.
  2005{\natexlab{b}}, \prd, 71, 114011

\bibitem[{{Alford} {et~al.}(2003){Alford}, {Bowers}, {Cheyne}, \&
  {Cowan}}]{2003PhRvD..67e4018A}
{Alford}, M.~G., {Bowers}, J.~A., {Cheyne}, J.~M., \& {Cowan}, G.~A. 2003,
  \prd, 67, 054018

\bibitem[{{Anglani} {et~al.}(2013){Anglani}, {Casalbuoni}, {Ciminale}, {Gatto},
  {Ippolito}, {Mannarelli}, \& {Ruggieri}}]{2013arXiv1302.4264A}
{Anglani}, R., {Casalbuoni}, R., {Ciminale}, M., {et~al.} 2013, ArXiv e-prints
  1302.4264

\bibitem[{{Anglani} {et~al.}(2006){Anglani}, {Nardulli}, {Ruggieri}, \&
  {Mannarelli}}]{2006PhRvD..74g4005A}
{Anglani}, R., {Nardulli}, G., {Ruggieri}, M., \& {Mannarelli}, M. 2006, \prd,
  74, 074005

\bibitem[{{Antoniadis} {et~al.}(2013){Antoniadis}, {Freire}, {Wex}, {Tauris},
  {Lynch}, {van Kerkwijk}, {Kramer}, {Bassa}, {Dhillon}, {Driebe}, {Hessels},
  {Kaspi}, {Kondratiev}, {Langer}, {Marsh}, {McLaughlin}, {Pennucci}, {Ransom},
  {Stairs}, {van Leeuwen}, {Verbiest}, \& {Whelan}}]{2013arXiv1304.6875A}
{Antoniadis}, J., {Freire}, P.~C.~C., {Wex}, N., {et~al.} 2013, Science, 340,
  6131

\bibitem[{{Blaschke} {et~al.}(2012){Blaschke}, {Grigorian}, {Voskresensky}, \&
  {Weber}}]{2012PhRvC..85b2802B}
{Blaschke}, D., {Grigorian}, H., {Voskresensky}, D.~N., \& {Weber}, F. 2012,
  \prc, 85, 022802

\bibitem[{{Bonanno} \& {Sedrakian}(2012)}]{2012A&A...539A..16B}
{Bonanno}, L. \& {Sedrakian}, A. 2012, \aap, 539, A16

\bibitem[{{Casalbuoni} {et~al.}(2005){Casalbuoni}, {Gatto}, {Ippolito},
  {Nardulli}, \& {Ruggieri}}]{2005PhLB..627...89C}
{Casalbuoni}, R., {Gatto}, R., {Ippolito}, N., {Nardulli}, G., \& {Ruggieri},
  M. 2005, Physics Letters B, 627, 89

\bibitem[{{Demorest} {et~al.}(2010){Demorest}, {Pennucci}, {Ransom}, {Roberts},
  \& {Hessels}}]{2010Natur.467.1081D}
{Demorest}, P.~B., {Pennucci}, T., {Ransom}, S.~M., {Roberts}, M.~S.~E., \&
  {Hessels}, J.~W.~T. 2010, \nat, 467, 1081

\bibitem[{{Gatto} \& {Ruggieri}(2007)}]{2007PhRvD..75k4004G}
{Gatto}, R. \& {Ruggieri}, M. 2007, \prd, 75, 114004

\bibitem[{{Giannakis} {et~al.}(2005){Giannakis}, {Hou}, \&
  {Ren}}]{2005PhLB..631...16G}
{Giannakis}, I., {Hou}, D.-F., \& {Ren}, H.-C. 2005, Physics Letters B, 631, 16

\bibitem[{{Glen} \& {Sutherland}(1980)}]{1980ApJ...239..671G}
{Glen}, G. \& {Sutherland}, P. 1980, \apj, 239, 671

\bibitem[{{Grigorian} {et~al.}(2005){Grigorian}, {Blaschke}, \&
  {Voskresensky}}]{2005PhRvC..71d5801G}
{Grigorian}, H., {Blaschke}, D., \& {Voskresensky}, D. 2005, \prc, 71, 045801

\bibitem[{{Gudmundsson} {et~al.}(1983){Gudmundsson}, {Pethick}, \&
  {Epstein}}]{1983ApJ...272..286G}
{Gudmundsson}, E.~H., {Pethick}, C.~J., \& {Epstein}, R.~I. 1983, \apj, 272,
  286

\bibitem[{{Heinke} \& {Ho}(2010)}]{2010ApJ...719L.167H}
{Heinke}, C.~O. \& {Ho}, W.~C.~G. 2010, \apjl, 719, L167

\bibitem[{{Hess} \& {Sedrakian}(2011)}]{2011PhRvD..84f3015H}
{Hess}, D. \& {Sedrakian}, A. 2011, \prd, 84, 063015

\bibitem[{{Huang} \& {Sedrakian}(2010)}]{2010PhRvD..82d5029H}
{Huang}, X.-G. \& {Sedrakian}, A. 2010, \prd, 82, 045029

\bibitem[{{Ippolito} {et~al.}(2008){Ippolito}, {Ruggieri}, {Rischke},
  {Sedrakian}, \& {Weber}}]{2008PhRvD..77b3004I}
{Ippolito}, N.~D., {Ruggieri}, M., {Rischke}, D.~H., {Sedrakian}, A., \&
  {Weber}, F. 2008, \prd, 77, 023004

\bibitem[{{Iwamoto}(1980)}]{1980PhRvL..44.1637I}
{Iwamoto}, N. 1980, Physical Review Letters, 44, 1637

\bibitem[{{Jaikumar} {et~al.}(2006){Jaikumar}, {Roberts}, \&
  {Sedrakian}}]{2006PhRvC..73d2801J}
{Jaikumar}, P., {Roberts}, C.~D., \& {Sedrakian}, A. 2006, \prc, 73, 042801

\bibitem[{{Kl{\"a}hn} {et~al.}(2007){Kl{\"a}hn}, {Blaschke}, {Sandin}, {Fuchs},
  {Faessler}, {Grigorian}, {R{\"o}pke}, \& {Tr{\"u}mper}}]{2007PhLB..654..170K}
{Kl{\"a}hn}, T., {Blaschke}, D., {Sandin}, F., {et~al.} 2007, Physics Letters
  B, 654, 170

\bibitem[{{Knippel} \& {Sedrakian}(2009)}]{2009PhRvD..79h3007K}
{Knippel}, B. \& {Sedrakian}, A. 2009, \prd, 79, 083007

\bibitem[{{Kurkela} {et~al.}(2010){Kurkela}, {Romatschke}, {Vuorinen}, \&
  {Wu}}]{2010arXiv1006.4062K}
{Kurkela}, A., {Romatschke}, P., {Vuorinen}, A., \& {Wu}, B. 2010, ArXiv
  e-prints 1006.4062

\bibitem[{{Mannarelli} {et~al.}(2007){Mannarelli}, {Rajagopal}, \&
  {Sharma}}]{2007PhRvD..76g4026M}
{Mannarelli}, M., {Rajagopal}, K., \& {Sharma}, R. 2007, \prd, 76, 074026

\bibitem[{{M{\"u}ther} \& {Sedrakian}(2003)}]{2003PhRvD..67h5024M}
{M{\"u}ther}, H. \& {Sedrakian}, A. 2003, \prd, 67, 085024

\bibitem[{{Negreiros} {et~al.}(2013){Negreiros}, {Schramm}, \&
  {Weber}}]{2013PhLB..718.1176N}
{Negreiros}, R., {Schramm}, S., \& {Weber}, F. 2013, Physics Letters B, 718,
  1176

\bibitem[{{Page} {et~al.}(2000){Page}, {Prakash}, {Lattimer}, \&
  {Steiner}}]{2000PhRvL..85.2048P}
{Page}, D., {Prakash}, M., {Lattimer}, J.~M., \& {Steiner}, A.~W. 2000,
  Physical Review Letters, 85, 2048

\bibitem[{{Page} {et~al.}(2011){Page}, {Prakash}, {Lattimer}, \&
  {Steiner}}]{2011PhRvL.106h1101P}
{Page}, D., {Prakash}, M., {Lattimer}, J.~M., \& {Steiner}, A.~W. 2011,
  Physical Review Letters, 106, 081101

\bibitem[{{Potekhin} {et~al.}(1997){Potekhin}, {Chabrier}, \&
  {Yakovlev}}]{1997A&A...323..415P}
{Potekhin}, A.~Y., {Chabrier}, G., \& {Yakovlev}, D.~G. 1997, \aap, 323, 415

\bibitem[{{Rajagopal} \& {Sharma}(2006)}]{2006PhRvD..74i4019R}
{Rajagopal}, K. \& {Sharma}, R. 2006, \prd, 74, 094019

\bibitem[{{Schmitt}(2005)}]{2005PhRvD..71e4016S}
{Schmitt}, A. 2005, \prd, 71, 054016

\bibitem[{{Schmitt} {et~al.}(2003){Schmitt}, {Wang}, \&
  {Rischke}}]{2003PhRvL..91x2301S}
{Schmitt}, A., {Wang}, Q., \& {Rischke}, D.~H. 2003, Physical Review Letters,
  91, 242301

\bibitem[{{Sedrakian}(2013)}]{2013arXiv1301.2675S}
{Sedrakian}, A. 2013, PoS (Confinment X), 251

\bibitem[{{Sedrakian} \& {Rischke}(2009)}]{2009PhRvD..80g4022S}
{Sedrakian}, A. \& {Rischke}, D.~H. 2009, \prd, 80, 074022

\bibitem[{{Shovkovy} \& {Huang}(2003)}]{2003PhLB..564..205S}
{Shovkovy}, I. \& {Huang}, M. 2003, Physics Letters B, 564, 205

\bibitem[{{Shovkovy} {et~al.}(2005){Shovkovy}, {R{\"u}ster}, \&
  {Rischke}}]{2005JPhG...31S.849S}
{Shovkovy}, I.~A., {R{\"u}ster}, S.~B., \& {Rischke}, D.~H. 2005, Journal of
  Physics G Nuclear Physics, 31, 849

\bibitem[{{Shternin} {et~al.}(2011){Shternin}, {Yakovlev}, {Heinke}, {Ho}, \&
  {Patnaude}}]{2011MNRAS.412L.108S}
{Shternin}, P.~S., {Yakovlev}, D.~G., {Heinke}, C.~O., {Ho}, W.~C.~G., \&
  {Patnaude}, D.~J. 2011, \mnras, 412, L108

\end{thebibliography}

\end{document}